\documentclass[11pt,letterpaper]{article}
\usepackage[T1]{fontenc}
\usepackage{graphics}
\usepackage[dvips]{epsfig}
\usepackage{eepic}
\usepackage{color}
\usepackage{graphicx}
\usepackage{amssymb,amsbsy,amsmath,amsfonts,amssymb,amscd}
\usepackage{setspace}

\usepackage[round,colon]{natbib}
\renewcommand{\v}[1]{\underline{#1}{}}
\newcommand{\vv}[1]{\boldsymbol{#1}}

\newcommand{\vu}{\v{u}}
\newcommand{\vx}{\v{x}}

\newcommand{\cauchy}{\vv{\sigma}}

\newcommand{\Cauchy}{\vv{\Sigma}}

\renewcommand{\d}{\vv{d}}
\newcommand{\D}{\vv{D}}

\newcommand{\s}{\vv{s}}

\newcommand{\tun}{\vv{\delta}}
\newcommand{\Oml}{\Omega_{\ell}}
\newcommand{\Oms}{\Omega_{s}}
\newcommand{\dOm}{\partial \Omega}

\newcommand{\dtilde}{\tilde d}
\newcommand{\Dtilde}{\tilde D}
\newcommand{\dtildeeff}{{\tilde d}^{\text{eff}}}

\newcommand{\etahom}{\eta^{\text{hom}}}

\newcommand{\muhom}{\mu^{\text{hom}}}
\newcommand{\mul}{\mu_{\ell}}
\newcommand{\musct}{\mu^{\text{sct}}}

\newcommand{\tauc}{\tau_{c}}
\newcommand{\tauchom}{\tau_{c}^{\text{hom}}}

\newcommand{\dans}[1]{( \forall \; \v{x} \in {#1} )}

\newcommand{\gradvv}{{{\mbox{\bf grad}}\:}}
\newcommand{\refe}[1]{(\ref{#1})}

\newcommand{\tr}{\mbox{tr}}

\newcommand{\Reel}{\mbox{I\kern-2ptR}}
\newcommand{\Naturel}{\mbox{I\kern-1ptN}}
\newcommand{\transp}[1]{{}^{\mbox{t}}\!{#1}}
\newcommand{\dex}{\left (\vx \right)}
\catcode`@=11
\def\monequation{
    \bgroup
    \def\@@eqncr{\let\@tempa\relax 
        \ifcase\@eqcnt \def\@tempa{& &}\or \def\@tempa{&} 
        \else \def\@tempa{}\fi 
        \@tempa \if@eqnsw\@eqnnum\stepcounter{equation}\fi
        \global\@eqnswtrue\global\@eqcnt\z@\cr}
    \stepcounter{equation}\let\@currentlabel=\theequation
    \global\@eqnswtrue
    \global\@eqcnt\z@\let\\=\@eqncr
    $$\halign to \textwidth\bgroup\@eqnsel
    \tabskip\@centering$\displaystyle{##}$
    &\global\@eqcnt\@ne 
    \hfil$\displaystyle{##}$\hfil
    &\llap{##}\tabskip\z@\cr}
\def\endmonequation{\@@eqncr\egroup
    \global\advance\c@equation\m@ne$$\egroup\global\@ignoretrue}
\catcode`@=12
\title{Homogenization approach to the behavior of suspensions of
  noncolloidal particles in yield  stress fluids}

\author{Xavier Chateau\footnote{corresponding author:
    xavier.chateau@lcpc.fr} ,
Guillaume Ovarlez, Kien Luu Trung \\
Laboratoire des Mat\'eriaux et des Structures du G\'enie Civil \\ 
Institut Navier (UMR113 LCPC-ENPC-CNRS) \\
  2 All\'ee Kepler, 77420 Champs sur Marne, France.}
\date{}
\begin{document}
%
\maketitle

\renewcommand{\abstractname}{Synopsis}
\begin{abstract}
The behavior of suspensions of rigid particles in a non-Newtonian
fluid is studied in the framework of a nonlinear
homogenization method. 
Estimates for the overall properties of the composite material are
obtained.
In the case of a Herschel-Bulkley suspending fluid, it is shown that
the properties of a suspension with overall isotropy can be
satisfactory  modeled as that of a Herschel-Bulkley fluid with an
exponent equal to that of the suspending fluid.  
Estimates for the yield stress and the consistency at large
strain rate levels are proposed.
These estimates compare well to both experimental data obtained
by~\cite{Mahaut-Chateau-Coussot-Ovarlez-2007} and to experimental data
found in the literature.
\end{abstract}

\section{Introduction}
\label{sec:Introduction}

Heterogeneous systems consisting of particles suspended in a fluid
medium make up in a wide variety of materials of practical interest,
both natural (slurries, debris flows, lavas, ...) or
man-made (concretes, food pastes, paints, cosmetics, ...).
This abundance explains why the behavior of these materials have been
extensively studied from both a theoretical and an experimental point
of view.

In the framework of man-made materials, it does not seem to exist a
well established method to obtain  a material with given
rheological characteristics from components having known properties
despite the fact that is is a problem encountered in numerous industrial
processes.
The set-up of such a methodology requires the ability to
predict the overall behavior of the material from that of its
constituents.
A general answer for this problem has yet to be
found.

The main difficulty in modeling the behavior of
suspensions comes from the fact that the material is multiscale and
contains many interacting constituents. 
Experimental data show that the overall rheological properties of a
suspension depend upon the shape and size of the particles, the
interaction between particles (colloidal or noncolloidal), the
interaction between the particles and the suspending fluid (hydrodynamic), 
the properties of the suspending fluid (Newtonian or not) and the type
of flow the suspension is subjected to.

In this paper, we focus on suspensions made up of noncolloidal
particles dispersed in a yield stress fluid such as suspensions made
up of a coarse fraction and a colloidal fraction.
If the colloidal particles are much smaller than the coarse ones, the
latter interact with the other components of the suspension
only through hydrodynamic interactions: they see a homogeneous phase fluid
which behavior is that of the colloidal
suspension~[\citet{Sengun1989a,Sengun1989b,Ancey-Jorrot-2001}].

Like for any other suspensions, the rheological properties of non-Newtonian
suspensions depend upon the shape, the surface
texture and the size distribution of the coarse particles.

In addition to experimental approaches, modeling the behavior of
non-Newtonian suspensions using their constituents characteristics in
the framework of numerical methods appears promising because of the
ability of these methods to track the particles localization in the
suspension and to account for several interparticle interactions.
Solving the problem for noncolloidal particles immersed in a
Newtonian fluid is feasible using modern computers and provides useful
results~[\cite{Brady-2001}].
The most serious drawback of these methods to their
generalization to non-Newtonian suspending fluid is the fact that the
Newtonian hydrodynamic force applied to the particle is evaluated by means of
close form equations depending on the macroscopic strain rate 
and on the particle velocity relative to the bulk
fluid~[\cite{Bossis84,Brady-1988}].
Thanks to these relations, it is not necessary to solve the
continuous Stokes equation in the bulk fluid in order to simulate the
behavior of the suspension.
This enables to simulate the flow of
representative elementary volume of a suspension containing numerous
particles at a reasonable computational cost.
To our knowledge, such an equation does not exist for particles immersed
in a yield stress fluid, which prevents a generalization of the method
to be easily proposed.

Trying to evaluate the velocity field in the bulk yield
stress fluid by means of a finite element method is such a
time consuming task that no situation of practical interest can be
solved using these tools
today~[\cite{Johnson-Tezduyar1997,Roquet2003,Yu-Wachs2007}].

To obtain estimates for the overall characteristics of non-Newtonian
suspensions, the change of scale method appears then to be a powerful
tool.
As a reminder homogenization technique aims at identifying the
macroscopic properties of a material modeled as a continuous medium
from those of their constituents.
The first results in this field were obtained by~\cite{Einstein-1906}
for the viscosity of a dilute suspension.
Since, different problems have been studied such as the
viscosity of multimodal suspensions [\cite{Farris68}], the viscosity
of concentrated suspensions [\cite{Krieger59,Frankel-Acrivos-1967}],
the effect of the shape and orientation of particles on the
behavior of Newtonian suspensions~[\cite{Batchelor-1971}],
the effect of interparticle interactions~[\cite{Batchelor-1972}] or
the effect of Brownian
motions on the overall
behavior~[\cite{Batchelor-1977,Russel-Saville-Schowalter-1995}].

It is worth noticing that it  has not been possible to obtain exact
solutions for most of the problems cited above; generally, only estimates
of the overall characteristics of the suspension have been obtained.
This situation is very similar to the one prevailing in the field of
homogenization approaches to the behavior of solid
materials~[\cite{Zaoui02}].
This is not surprising since the two
problems are very similar as it has been recognized
by~\cite{Batchelor-1972}. 
The main difference comes from the fact that the morphology of the
heterogeneities within the representative elementary volume is a given
for solid materials whereas it is an unknown for
particle suspensions since the flow of the
representative elementary volume of suspension and the morphology of
the particles are coupled

Interestingly, novel developments have taken place in  nonlinear
continuum micromechanics in the last twenty years so that  both
estimates and variational bounds are now
available~[see \cite{Ponte-Castaneda-Suquet-1998} for a review].
Because of the similarities between the solid problem with the liquid
one, it is quite natural to address the modeling of the behavior of
non-Newtonian suspensions within this framework.

The aim of this paper is to provide a first approach to the overall
behavior of a suspension of non-Brownian and noncolloidal particles
immersed in an incompressible yield stress
fluid.
First, the main features of the homogenization approach to the behavior of
non-Newtonian suspensions are recalled.
Then, the secant-method of~\cite{Ponte-Castaneda-1991} and~\cite{Suquet-1993}
is applied in order to estimate the overall behavior of the suspension.
The estimates are compared to new experimental
results obtained by~\cite{Mahaut-Chateau-Coussot-Ovarlez-2007} and to
experimental results of the literature in the next part.
Finally, the validity of the theoretical approach is discussed before
some general conclusions are drawn.

\section{Homogenized behavior}
We examine the overall behavior of a suspension, the liquid phase
of which is homogeneous and nonlinear.
We restrict ourselves to the situation where the constitutive behavior
of the fluid is characterized by an energy function $w
\left(\d\right)$ where ${\tilde d}  = \sqrt{2 \d : \d}$ denotes the
second invariant of the Eulerian strain rate tensor $\d$.
The Cauchy stress $\cauchy$ is obtained by differentiation of the
potential $w$ with respect to the strain
rate tensor $\d$ if  $w$ is differentiable.
Otherwise, the derivative should be interpreted as
the subdifferential of convex analysis.

The condition of naught strain for the particles can also be written
in term of a dissipation potential with $w$ defined by
\begin{equation}
  \label{eq:Def-w-indeformable}
  w \left(\d\right) = 0 \text{ if } \d = 0 \quad   w \left(\d\right) =
  \infty \text{ if }  \d \neq 0
\end{equation}

Thus, the suspension is made of a heterogeneous medium, the behavior
of which is described by
\begin{equation}
  \label{eq:LdCgen}
  \cauchy = \frac{\partial w}{\partial \d}
  \left(\d, \vx \right)
\end{equation}
$w$ being a strict convex function.
The location in the representative elementary volume is defined by the
position vector $\vx$. 

\begin{figure}[htbp] 
\begin{center}
\input{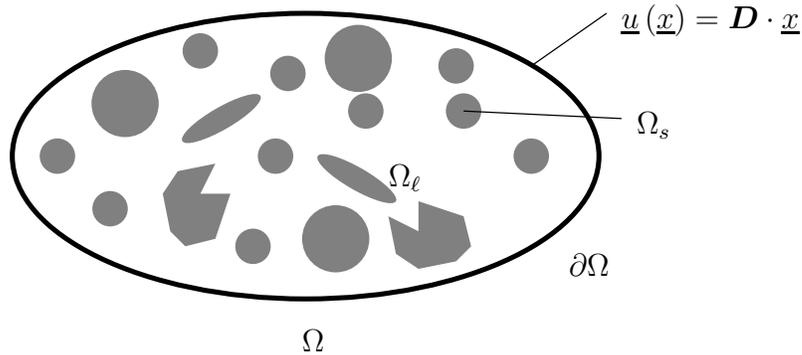}
\caption{The representative elementary volume of the
  suspension submitted to a macroscopic strain rate loading (Hashin
  boundary condition).
\label{fig:Hashin}}
\end{center} 
\end{figure}

It is assumed that it is possible to define a
representative elementary volume of the
suspension occupying a domain $\Omega$ with boundary $\partial
\Omega$ such that it is large enough to be of typical composition and
its overall properties do not depend on the way it is loaded at the
macroscopic scale.
$\Oms$ and $\Oml$ denote the solid and the liquid
domain respectively. 
The volume fraction of particles $\varphi$ is the ratio of the volume
fraction of $\Oms$ in $\Omega$. 
For simplicity, it is assumed that the boundary of $\Omega$ is located
in the liquid domain as depicted in Fig.~\ref{fig:Hashin}.
At the microscopic scale, $\Omega$ is considered as a structure. 
$<a>$ (resp. $<a>_\alpha$ with $\alpha=s,\,\ell$) denotes the average
of $a$ over $\Omega$ (resp. $\Omega_\alpha$).
The liquid phase is homogeneous. 
We adopt an Eulerian description of the movement and we restrict our
attention to the situations where the evolutions of the system are
quasistatics ({\em i.e.} inertial effects are negligible) and all the
long range forces other than the hydrodynamic ones are negligible.

As shown by~\cite{Hill-1963}, the overall behavior of the suspension reads:
\begin{equation}
  \label{eq:LdCMacro}
  \Cauchy = \frac{\partial W}{\partial \D} \left(\D\right) \quad
  \text{with } W \left(\D\right) = \min_{\d \in {\cal C}(\D)}
  \langle w \left(\d \right)\rangle
\end{equation}
where ${\cal C}\left(\D\right)$ denotes the set of Eulerian strain rate
fields kinematically admissible with $\D$.
It is recalled that a strain rate field $\d$ defined over $\Omega$ is
said to be kinematically admissible with $\D$ if exists a velocity
field $\vu$ defined over $\Omega$, complying with the Hashin
condition 
\begin{equation}
  \label{eq:Hashin}
  \dans{\dOm} \quad \vu \dex = \D \cdot \vx
\end{equation}
such as:
\begin{equation}
  \label{eq:def-d}
\dans{\Omega} \qquad 
\d \dex = {\displaystyle \frac{1}{2}  \left(\gradvv \vu
       \dex   + \transp \gradvv \vu \dex\right)} 
\end{equation}

When the fluid obeys a Herschel-Bulkley law
with a yield stress~$\tau_c$, a consistency~$\eta$ and a power law
exponent~$n > 0$, the dissipation potential reads:
\begin{equation}
  \label{eq:H-B-1}
  \begin{array}{lll}
    w \left(\d\right) = \tau_c \,{\tilde d} + {\displaystyle
      \frac{\eta}{n+1} {\tilde d}^{n+1}} & \text{if} & \tr \d = 0 \\
    \\
    w  \left(\d\right) = \infty & \text{if} & \tr \d \neq 0
  \end{array}
\end{equation}
Then the fluid's state equation reads 
\begin{equation}
  \label{eq:H-B-2}
  \begin{array}{llll}
 \d = 0 & & \text{if } & \sqrt{\s : \s/2} < \tau_c  \\
\\
\s = \left(\tau_c +  \eta
  {\tilde d}^{n}\right)  {\displaystyle\frac{\d}{\tilde d} } \quad
\tr{\d} = 0 & & \text{if} &\sqrt{\s : \s/2} \ge \tau_c 
  \end{array}
\end{equation}
where $p =  - \tr \cauchy/3$ is the hydrostatic pressure, $\s =
\cauchy + p \tun$ the deviator of $\cauchy$ and $\tun$ the second
order unit tensor.
As it is common for incompressible materials, the pressure $p$ is not
determined by the state law.
Considering that the particles are rigid and that the bearing
fluid is homogeneous and incompressible, it is easily shown from
Eq.~\refe{eq:LdCMacro} that the macroscopic potential
is also defined by
\begin{equation}
  \label{eq:LdCMacro2}
W \left(\D\right) = \min_{\d \in {\cal G}\left(\D\right)}
\left(1-\varphi\right)
  \left\lbrack  \tau_c  \left\langle{\tilde d} \right\rangle_{\ell} +
    \frac{\eta}{n+1}  \left\langle{\tilde d}^{n+1}
    \right\rangle_{\ell}\right\rbrack \qquad \text{if } \tr \D =0 
\end{equation}
where ${\cal G} \left( \D\right)$ denotes the subset of ${\cal C}
\left( \D\right)$ which elements comply with the naught strain rate
constraint over the domain occupied by the particles and the
incompressible constraint over the fluid domain.
Of course, the set ${\cal G}\left(\D\right)$ is only defined for
macroscopic strain rate complying with the condition $\tr{\D} = 0$. 
Eq.~\refe{eq:LdCMacro2} is completed by the condition $W
\left(\D\right) = \infty$ if $\tr \D \neq 0$, which enforced the
incompressible constraint at the macroscopic level.

The set ${\cal G}\left(\D\right)$  being convex, the minimization
problem~\refe{eq:LdCMacro2} admits only one solution, which ensures the
validity of the method.
The identification of the macroscopic behavior of the suspension from
Eq.~\refe{eq:LdCMacro} or Eq.~\refe{eq:LdCMacro2} requires the
resolution of a continuous convex minimization problem. 
This problem of minimization has to be solved for each morphology of
the suspension defined by the shape of the particles and the
distribution of the particles within the suspension. 
Of course, it is not possible to solve this problem in most situations
of practical interest. 

To remedy this difficulty, various estimation techniques of the
macroscopic behavior have been proposed, in particular,
by~\cite{Ponte-Castaneda-1991,Ponte-Castaneda-1996,Ponte-Castaneda-2003}
and \cite{Suquet-1993}.

The key feature in these methods is the use of a rigorous variational
principle (Eq.~\ref{eq:LdCMacro} for the yield stress
fluid suspensions) to determine the best possible choice of a ``linear
comparison composite'' to estimate the effective behavior of the
nonlinear one.
A detailed description of these methods is beyond the scope of the
present paper.
The reader is referred to~[\cite{Ponte-Castaneda-Suquet-1998}] for a
more detailed review.

In the following, we obtained estimates relevant to our problem
in the framework of a simple approach which does make
use explicitly of the variational Eq.~\ref{eq:LdCMacro}.
The main features of the method used to obtain these estimates are
recalled in the following section (largely inspired by the presentation
of~\cite{Suquet-1997}) for completeness of the paper.

\section{Secant estimate of the behavior}
\label{sec:secant}

It is assumed  that the solid particles are
isotropically distributed over the representative elementary volume.
The macroscopic behavior of the suspension is therefore also
isotropic.

\begin{figure}[htbp] 
\begin{center}
\input{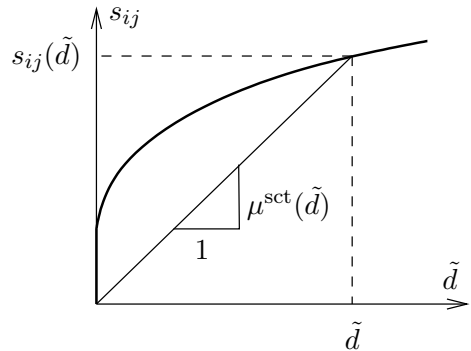}
\caption{Secant modulus formulation of nonlinear incompressible
  isotropic fluid.
\label{fig:Secantmodulus}}
\end{center} 
\end{figure}

It is possible to write the behavior of the nonlinear fluid in the following form 
\begin{equation}
  \label{eq:LdC-secante}
  \cauchy = 2 \musct ({\tilde d}) \d - p \tun
\end{equation}
where $\musct(\tilde
d)$ denotes the secant modulus of the fluid phase.
The secant modulus is no more than the
apparent viscosity of the fluid, as depicted in
Fig.~\ref{fig:Secantmodulus}.

Using Eq.~\refe{eq:LdC-secante}
for the state equation of the suspending fluid, it is possible to replace the
original nonlinear problem of homogenization by a linear
homogenization problem for a suspension of rigid particles immersed
in a heterogeneous incompressible isotropic fluid with
apparent viscosity~$\eta\dex =  \musct \left(\v{x}, {\tilde d} \dex\right)$.
 
Then, the difficulty which remains to be solved is to calculate, or to
estimate, the field ${\tilde d}$ over the domain occupied by the fluid
phase in the representative elementary volume.
Since it is impossible to analytically determine the local
response of the nonlinear fluid, it is impossible to compute the
secant modulus field over the representative elementary volume.
Then an approximation has to be introduced to make analytic
calculations feasible. 
The approximation for secant methods consists of replacing the secant
modulus field over the representative elementary volume by a modulus
which is uniform over subdomains of the representative elementary
volume.
In this paper,  we consider only one domain for
the liquid phase to simplify.
Then the secant modulus is uniform over the fluid phase. 
This estimate reads:
\begin{equation}
  \label{eq:mu-secant-phase}
  \dans{\Oml} \quad \musct  \left(\v{x}, \dtilde \dex \right) \simeq
  \musct_{\ell} (\dtildeeff_{\ell}) 
\end{equation}
where $\dtildeeff_{\ell}$ is an equivalent effective strain rate which
remains to be defined as a function of the mean value of the
field~$\dtilde$ in the fluid phase, and thus, of the value of the
macroscopic strain rate. 
The replacement of the field of heterogeneous secant modulus defined
over the fluid phase by a homogeneous field simplifies considerably
the resolution of the problem. 
Consequently, the estimate technique of the nonlinear macroscopic
behavior includes three steps:
\begin{enumerate}
\item First of all, it is necessary to solve a linear homogenization
  problem for a suspension of particles immersed in an isotropic
  homogeneous fluid of viscosity~$\mu_{\ell}$. 
  If the suspension is isotropic at the macroscopic scale, the
  macroscopic behavior is characterized by a macroscopic viscosity
  proportional to the viscosity of the fluid, the coefficient of
  proportionality depending on the morphology of the particles. 
  As this problem has been the subject of many works, numerous results
  and estimates are available in the literature dealing with the
  rheology of suspensions or the homogenization to the behavior of
  heterogeneous linear materials.
  Let $g(\varphi)$ denotes the coefficient of proportionality linking the
  macroscopic viscosity to the microscopic viscosity.
  \begin{equation}
    \label{eq:Def-g}
    \muhom = \mul \: g(\varphi)
  \end{equation}
\item Then, one must choose a measure of the effective strain rate for the
  fluid phase. 
  As the considered  material is isotropic at the microscopic
  scale, the strain rate is characterized by the second order moment
  of the quantity~$\dtilde$ defined by:
  \begin{equation}
    \label{eq:def-deffl}
    \dtildeeff_{\ell} = \sqrt{\langle\dtilde^2\rangle_{\ell}}
  \end{equation}
  This choice corresponds to the modified approach described
  by~\cite{Suquet-1997}. 
  \cite{Ponte-Castaneda-1991}
  demonstrated that this choice is optimal in the framework of a
  variational approach to the solution of the nonlinear homogenization
  problem under consideration.
  It would have been simpler to choose $\dtildeeff_{\ell} =
  \widetilde{<\d>}_{\ell}$, a quantity easily computed from the
  egality  $\left<\d \right>_{\ell} = \D / {(1 - \varphi})$.
  Unfortunately, \cite{Suquet-1997} showed  that estimates of the
  overall properties of the heterogeneous material obtained using
  this effective liquid strain rate are less accurate than those
  obtained using Eq.~\refe{eq:def-deffl}.
\item Finally, the nonlinear character of the problem is taken into
  account by integrating into the relation of homogenization~\refe{eq:Def-g}
  linking~$\muhom$ to the viscosity of the liquid phase, the fact that
  the value of the fluid viscosity depends on~$\dtildeeff_{\ell}$. 
  As~$\dtildeeff_{\ell}$ depends on the value of the macroscopic
  strain rate $\D$, the value of the macroscopic secant modulus also
  depends on the value of~$\D$.
  The only difficulty implementing this step is the calculation
  of~$\dtildeeff_{\ell}$ as a function of $\D$ for the particular
  homogenization scheme used. 
  It has been shown by~\cite{Kreher90} that:
  \begin{equation}
    \label{eq:Kreher} 
    \langle \dtilde^2 \rangle_{\ell} = \frac{1}{1-\varphi} \frac{\partial
    \muhom}{\partial \mul} \Dtilde^2 =  \frac{1}{1-\varphi}
  g(\varphi)  \Dtilde^2
  \end{equation}
\end{enumerate}
  By remplacing $\mul$ by $\musct$ in the relation of linear
  homogenization~Eq.~\refe{eq:Def-g} and then by combining the obtained
  equation with the localization Eq.~\refe{eq:Kreher},
  one obtains the following estimate for the  macroscopic secant
  modulus (i.e. apparent viscosity) of any nonlinear materials with
  any isotropic microstructure:
\begin{equation}
    \label{eq:muscthom}
   \muhom \left(\varphi, {\tilde D}\right) = g(\varphi)  * \musct
   \left({\tilde d}  \right) \quad \text{ with } \quad {\tilde d} =
 {\tilde D} \sqrt{\frac{g(\varphi)}{1-\varphi}} 
\end{equation}
Using notations of \cite{Sengun1989b},
the apparent viscosity Eqs.~\refe{eq:muscthom} reads 
\begin{equation}
  \label{eq:segun}
  \eta (\varphi, {\dot \gamma})=  \eta_{\text{cr}} (\varphi) *
  \eta_{\text{fr}} ({\dot \gamma}_{\text{eff}}) 
  \quad \text{ with } \quad {\dot \gamma}_{\text{eff}} =
  {\dot \gamma} \sqrt{\frac{\eta_{\text{cr}}(\varphi)}{1-\varphi}}  
\end{equation}
which is much more general
than  Eqs~(2.8), (2.9), (4.1) and (4.2) of~\cite{Sengun1989b} as they
allow to take into account any estimate $g(\varphi)$ ({\em i.e.}
$\eta_{\text{cr}} (\varphi)$) of the relative viscosity of a Newtonian
suspension.
It is worth noting that Eqs.~\refe{eq:muscthom} and \refe{eq:segun}
are valid for any particles shape and dispersity.
Interestingly, even if the estimate $g(\varphi)$ does not rely on a
morphological model, Eq.~\refe{eq:Kreher} allows to estimate the
localization factor associated with the relative viscosity
function~$g(\varphi)$.

Coming back to the Herschel-Bulkley suspension problem, the secant modulus
of the suspending fluid reads:
\begin{equation}
  \label{eq:Def-mu-secant}
  \musct \left(\v{x}, {\tilde d} \dex\right) = \frac{\tau_c}{{\tilde d}
    \dex} + \eta \left({\tilde d} \dex\right)^{n-1}
\end{equation}
Then, putting Eq.~\refe{eq:Def-mu-secant} into
Eq.~\refe{eq:muscthom} yields
\begin{equation}
  \label{eq:mu-sct-hom}
  \muhom \left(\varphi, \D\right) = \frac{\tauchom}{\Dtilde} + \etahom
  \Dtilde^{n-1}
\end{equation}
  with:
\begin{equation}
  \label{eq:tauchom}
  \tauchom= \tauc \sqrt{(1-\varphi)  g(\varphi)}
\end{equation}
  and:
\begin{equation}
  \label{eq:etahom}
  \etahom = \eta g(\varphi) \left\lbrack\frac{g(\varphi)}{1-\varphi}
  \right\rbrack^{\frac{n-1}{2}}
\end{equation}
It is therefore predicted that, at the macroscopic scale, the
suspension behaves as a Herschel-Bulkley fluid with same exponent as
that of the suspending fluid. 
Overall yield stress and macroscopic consistency are defined
by Eqs.~\refe{eq:tauchom} and~\refe{eq:etahom}.
This result does not depend on the scheme ({\em i.e.} $g(\varphi)$)
used to link the viscosity of the bearing fluid to the overall
viscosity of the suspension. 
The quality of the prediction depends thus only on the validity of the
assumption that the field~$\musct_{\ell} \left(\v{x}, \dtilde \dex
\right)$ can be estimated by the quantity~$\musct_{\ell}
(\dtildeeff_{\ell})$.

\section{Experimental validation}

An experimental procedure has been designed
by~\citet{Mahaut-Chateau-Coussot-Ovarlez-2007} which complies with the
assumptions made to obtain the theoretical results presented
above.
As this procedure is described in detail in the paper entitled {\em
 Yield stress and elastic modulus of suspensions of noncolloidal
 particles in yield stress fluids}, no further
details concerning the experimental work are given in this paper;
we restrict ourselves to comparisons between  experimental data and
theoretical predictions.
 
\subsection{Elastic modulus vs. yield stress}

The accuracy of the estimates obtained in the framework of the
theoretical approach presented above for the overall properties of the yield
stress suspension depends on the assumption made to take into
account the nonlinear behavior of the suspending fluid and on the
scheme used to estimate the overall linear behavior of the
suspension.
It is possible to check experimentally the validity of the assumption
made on the heterogeneities of the secant modulus over the liquid
domain irrespective of the errors induced by the choice of a
particular homogenization scheme.
For this, it is enough to remark that 
Eq.~\refe{eq:Def-g} enables to calculate the macroscopic
elastic modulus $G^{\text{hom}}$ of an isotropic suspension of
particles dispersed in an isotropic incompressible linear elastic matrix
whose shear modulus is equal to~$G$ (both problems pose exactly in
the same way provided that $\d$ and $\mul$ be identified
with the infinitesimal strain tensor and the
elastic shear modulus).  
As a consequence, it is possible to obtain a general
relationship between the dimensionless elastic modulus and the
dimensionless yield stress of a suspension of rigid particles
dispersed  in a yield stress fluid  that is true whatever
the scheme as long as the particle distribution is isotropic and using an
uniform secant modulus estimate is relevant.
Combining Eqs.~\refe{eq:Def-g},\refe{eq:tauchom} and
\refe{eq:etahom} yields the relations:
\begin{equation}
  \label{eq:verifexp-tauc}
  {\tauchom}/{\tauc} = \sqrt{(1-\varphi)
  {G^{\text{hom}}}/{G}}
\end{equation}
and
\begin{equation}
  \label{eq:verifexp-eta}
  \etahom/\eta = \sqrt{\frac{\left(G^{\text{hom}} /
        G\right)^{n+1}}{\left(1-\varphi\right)^{n-1}}}
\end{equation}
Moreover, the yield stress and the consistency not being independent of one
another, it is also possible to determine the consistency from
the yield stress and the concentration:
\begin{equation}
  \label{eq:etahom2}
  \etahom/\eta =
  \frac{\left(\tauchom/\tauc\right)^{n+1}}{\left(1-\varphi\right)^{n}}
\end{equation}

\begin{figure}[htbp]
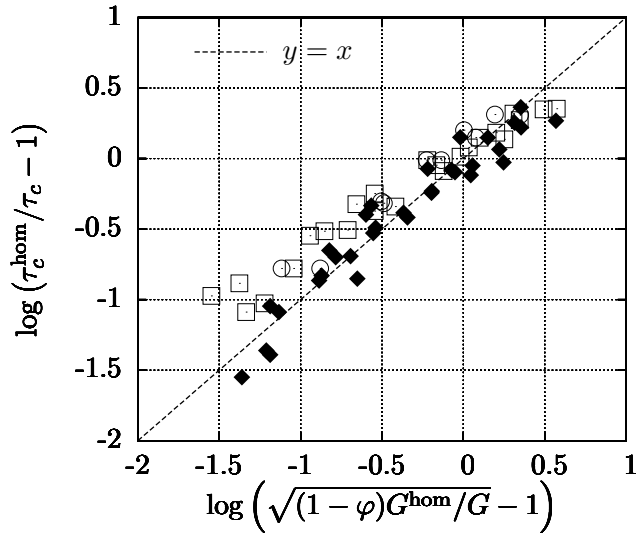

 	\centering
		\input figure10
\caption{Dimensionless yield stress $\tauchom/\tau_c$ as a
  function of  $\sqrt{(1-\varphi)G^{\text{hom}}/G}$ for
   all the systems studied by~\citet{Mahaut-Chateau-Coussot-Ovarlez-2007}.
   The open square symbols are suspensions of polystyrene and glass
   beads in bentonite, the open circle symbols are suspensions of glass
   beads in carbopol while the solid diamond symbols are suspensions of
   polystyrene and glass beads in emulsion.
   The figure's coordinates  were chosen so that the $y=x$ line
   represents the theoretical relation~\refe{eq:verifexp-tauc}.
   \label{elasticity_vs_yield_stress}}
\end{figure}

In Fig.~\ref{elasticity_vs_yield_stress}, we have plotted the
dimensionless yield stress $\tau_c/\tau_c$ as  a function
of the dimensionless quantity
$\sqrt{(1-\varphi)G^{\text{hom}}/G}$ for all the systems
studied by~\citet{Mahaut-Chateau-Coussot-Ovarlez-2007}.
It is recalled that yield stress fluids have a solid linear
viscoelastic behavior below the yield stress, so that the macroscopic
elastic modulus of the suspensions could be experimentally measured
through oscillatory shear measurements.
A good agreement between the experimental results and the
micromechanical estimation \refe{eq:verifexp-tauc} (which is plotted as
a straight line $y=x$ in these coordinates) is observed.
These results show that the data are consistent with the assumption
that an uniform estimate of the secant modulus over the fluid domain
allows to accurately estimate the overall properties of the
suspension in the studied situations.

\subsection{Elastic modulus}

In this section, we summarize the results of the elastic modulus
measurements performed on all the materials. 
\begin{figure}[htbp]
  \centering
	\input figure12 
  \caption{Dimensionless elastic modulus $G^{\text{hom}}/G$ vs the
    beads volume fraction $\varphi$ for all the systems studied
    by~\citet{Mahaut-Chateau-Coussot-Ovarlez-2007}. 
    The open square symbols are suspensions of polystyrene and glass
    beads in bentonite, the open circle symbols are suspensions of
    glass beads in carbopol while the solid diamond symbols are  suspensions of
    polystyrene and glass beads in emulsion. 
    The solid line is the Krieger-Dougherty Eq.~\refe{eq:K-D}
    with $\varphi_m = 0.57$.
	\label{fig:icmf2}}
\end{figure} 

The evolution of the dimensionless modulus $G^{\text{hom}}/G$ as a
function of the volume fraction $\varphi$ of noncolloidal particles
for all the studied materials are summarized in Fig.~\ref{fig:icmf2}.
It is  observed that the experimental data are very well fitted to the
Krieger-Dougherty law (\citet{Krieger59}):
\begin{equation}
  \label{eq:K-D}
  \frac{G^{\text{hom}}}{G} = g(\varphi) = 
  \left(1-\frac{\varphi}{\varphi_m}\right)^{-2.5 \varphi_m}
\end{equation}
The value of the maximum packing fraction $\varphi_m=0.57$ was fixed by
means of a least squares method.
This value is very close to the value $\varphi_m=0.605$ measured locally
very recently in dense suspensions of noncolloidal particles in
Newtonian fluids by \citet{Ovarlez06} through MRI techniques.
The small discrepancy between the two
values of $\varphi_m$ comes certainly from the anisotropy
induced by the flow in the experiments of~\citet{Ovarlez06} as it was
also observed in the experiments
performed by~\citet{Parsi-Galada-Maria-1987}.

\subsection{Yield stress}

We now present the results of the yield stress
measurements. 

The yield stress $\tau_c$ was measured with a  method which avoids
destroying the homogeneity and the isotropy of the suspension.
The evolution of the dimensionless yield stress $\tauchom/\tau_c$ as a
function of the solid volume fraction $\varphi$ of noncolloidal
particles is depicted in Fig.~\ref{fig:icmf1} for the studied materials.

\begin{figure}[htbp]
	\centering
		\input figure11
	\caption{Dimensionless yield stress $\tauchom/\tauc$ vs
   the beads volume fraction $\varphi$ for all the systems studied
   by~\citet{Mahaut-Chateau-Coussot-Ovarlez-2007}.
   The open square symbols are suspensions of polystyrene and glass
   beads in bentonite, the open circle symbols are suspensions of glass
   beads in carbopol while the solid diamond symbols are suspensions of
   polystyrene and glass beads in emulsion. 
          The solid line is the theoretical
          prediction~\refe{eq:taucKD} with $\varphi_m = 0.57$. 
          The dashed curve is the dilute estimate~\refe{eq:taucDL} for
          the yield stress.
	\label{fig:icmf1}}
\end{figure} 

Fig.~\ref{fig:icmf1} clearly shows that the yield stress of the
suspensions reads as the product of the
suspending fluid yield stress times a function $f(\varphi)$.
\begin{figure}[htbp]
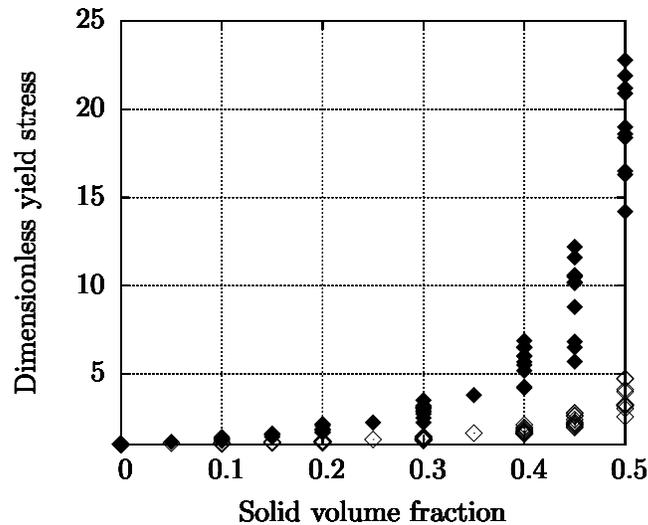

	\centering
		\input figure13
   \caption{Dimensionless yield stress $\tauchom/\tauc$ (empty
     symbols) and dimensionless elastic modulus $G^{\text{hom}}/G$
     (solid symbols) vs the beads volume fraction $\varphi$ for all
     the suspensions studied by~\citet{Mahaut-Chateau-Coussot-Ovarlez-2007}.
	\label{fig:seuilvselasticite}}
\end{figure} 
When both the dimensionless yield stresses and the dimensionless elastic
moduli are drawn on the same diagram (see
Fig.~\ref{fig:seuilvselasticite}), it is obvious that the yield stress
function $f(\varphi)$ is different from the elastic modulus function
$g(\varphi)$.

It is possible to directly use Eqs.~\refe{eq:mu-sct-hom},
\refe{eq:tauchom} and~\refe{eq:etahom} to evaluate the function
$f(\varphi)$.

For the dilute suspensions, the Einstein relation
$g^{\text{DL}}(\varphi) = 1 + 5/2 \varphi$ is exact to the first order 
of $\varphi$. 
Introducing this relation in Eqs. \refe{eq:tauchom}
and~\refe{eq:etahom} and keeping only the first order terms yield the
dilute estimates:
\begin{eqnarray}
  \label{eq:taucDL}
  \tauc^{\text{DL}} & = & \tauc \left(1 + 3/4 \varphi\right) \\
\nonumber\\ 
  \label{eq:etaDL}
  \eta^{\text{DL}} & =  &{\displaystyle \eta \left(1 +  \frac{7 n +
        3}{4} \varphi\right)}
\end{eqnarray}
The coefficients of growth of these two
estimates with the solid volume fraction are different from that of
the Einstein law. 
According to the experimental results depicted in
Fig.~\ref{fig:seuilvselasticite} the change of scale method also
predicts that the two functions $g(\varphi)$ and $f(\varphi)$ are
different.
The only approximation performed to obtain
Eqs.~\refe{eq:taucDL} and \refe{eq:etaDL} is to assume that a uniform
estimate of the secant modulus over the fluid domain enables to
accurately estimate the overall properties of the suspension.
Thus, estimates~\refe{eq:taucDL} and~\refe{eq:etaDL} are not rigorously
exact in contrast with the Einstein law.
It is worth noticing that for a suspension of particles in a Bingham
fluid, the overall consistency of the suspension is given by the
classical Einstein function $1+5/2 \varphi$.

For larger values of the solid volume fraction, the viscosity of a Newtonian
suspension is classically estimated using the Krieger-Dougherty
equation~[\citet{Krieger59,Quemada-1985}].
Putting the second equality of Eq.~\refe{eq:K-D} into Eqs.~\refe{eq:tauchom}
and~\refe{eq:etahom} yields the Krieger-Dougherty estimates for the
effective yield stress and the effective consistency of the
suspension:
\begin{eqnarray}
  \label{eq:taucKD}
  \tauc^{\text{KD}} & = & \tauc
  \sqrt{(1-\varphi)(1-\varphi/\varphi_m)^{-2.5 \varphi_m}}\\
\nonumber\\ 
  \label{eq:etaKD}
  \eta^{\text{KD}} & = & {\displaystyle \eta \left(1
      -\varphi\right)^{\frac{1-n}{2}}
    \left(1-\varphi/\varphi_m\right)^{-1.25(n+1) \varphi_m}}
\end{eqnarray}
Both the macroscopic yield stress and the
macroscopic consistency diverge when~$\varphi$ tends
towards~$\varphi_m$.
The values obtained with Eq.~\refe{eq:taucKD} are plotted in
Fig.~\ref{fig:icmf1}, taking $\varphi_m = 0.57$.
Good accordance is observed between theoretical and experimental data,
accounting for the validity of Eq.~\refe{eq:taucKD}.
Note that no attempt to fit the value of $\varphi_m$ has been performed.

\subsection{Experimental results from the literature}

In this paragraph, the theoretical estimates are compared with
experimental data of~\cite{Ancey-Jorrot-2001}, \cite{Erdogan-2005} and
\cite{Geiker-2002}.

The results for the yield stress are depicted in
Fig.~\ref{fig:seuil-biblio}.
The experimental data of~\citet{Geiker-2002} have not been drawn in
Fig.~\ref{fig:seuil-biblio} because they report to measure yield stress
ten times larger than other researchers.

\begin{figure}[htbp]
	\centering
		\input figure-seuil-biblio
	\caption{Dimensionless yield stress $\tauchom/\tauc$
           vs beads volume
          fraction $\varphi$ for suspensions
          of~\citet{Ancey-Jorrot-2001} (empty diamond-shaped)
          and~\citet{Erdogan-2005} (solid diamond-shaped).  
          The solid line is the theoretical
          prediction~\refe{eq:taucKD} with  $\varphi_m=0.635$ and the
          straight dashed line is the dilute overall yield
          stress~\refe{eq:taucDL}. The dashed curves are the
          theoretical Eq.~\refe{eq:taucKD} with $\varphi_m=0.67$
          (dashed curve below the solid curve) and $\varphi_m=0.6$
          (dashed curve above the solid curve).
	\label{fig:seuil-biblio}}
\end{figure} 

The theoretical curve in Fig.~\ref{fig:seuil-biblio} is calculated
using $\varphi_m = 0.635$, as reported by~\cite{Ancey-Jorrot-2001}.
We also plotted the curve using $\varphi_m = 0.67$ (dashed curve below
the solid curve) and $\varphi_m = 0.6$ (dashed curve above the solid
curve).
The experimental data of~\citet{Ancey-Jorrot-2001} are roughly well
fitted by the theoretical estimate~\refe{eq:taucKD} with $\varphi_m =
0.635$ even if the theoretical model always predicts yield stress
lower than the measured ones.
These discrepancies can not be impart to the experimental procedure.
The suspensions was prepared by adding monodisperse glass
beads to a clay dispersion and the yield stress was measured by means
of a slump  test, an experimental procedure which ensures that the
materials remain isotropic and homogeneous.
It is believed that the discrepancy between the experimental data
of~\cite{Ancey-Jorrot-2001} and our theoretical predictions could come
from a wrong evaluation of the experimental maximum packing fraction.
To support this opinion, it is worth noting that the estimate
$\varphi_m = 0.635$ of~\cite{Ancey-Jorrot-2001} was neither measured
nor estimated from experimental data.
Moreover, this value $0.635$ is greater than the values
reported in the literature for a dense suspension of non colloidal
suspension of monodisperse spherical particles [\cite{Ovarlez06}].
It can be seen in Fig.~\ref{fig:seuil-biblio} that the experimental
data are well fitted by Eq.~\refe{eq:taucKD} with $\varphi_m=0.6$.

\cite{Erdogan-2005} used a Couette-vane rheometer and applied
a strong preshear to the samples before to measure the overall
properties of the concrete suspension. 
Nevertheless,  his results compare rather well
to the theoretical ones, even if the lack of experimental data for 
values of the solid fraction larger than $0.4$ does not allow to draw
a definitve conclusion.

\begin{figure}[htbp]
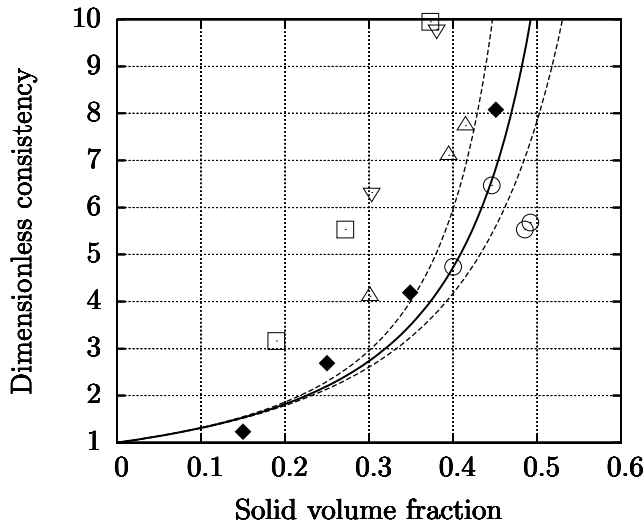

	\centering
		\input figure-viscosite-biblio
	\caption{Dimensionless consistency   $\etahom/\eta$ vs.
the solid volume fraction $\varphi$ for suspensions
of~\citet{Erdogan-2005} (solid diamond-shaped) and suspensions
of~\citet{Geiker-2002} (empty circle for glass beads, other
empty symbols for sea dredged, crushed and mixed aggregates). 
The solid line is the Krieger-Dougherty estimate for the consistency
($n=1$) with $\varphi_m=0.65$. The dashed lines are the same estimate
with $\varphi_m=0.55$ (above the solid curve) and $\varphi_m=0.75$
(below the solid curve).
	\label{fig:viscosite-biblio}}
\end{figure}

We now pay attention to the results of the consistency measurements.
Dimensionless consistency $\etahom/\eta$ measured by~\citet{Erdogan-2005}
and~\cite{Geiker-2002}  are plotted in
Fig.~\ref{fig:viscosite-biblio} as a function of the normalized solid
 volume fraction $\varphi/\varphi_m$.
Four types of aggregates were used
by~\cite{Geiker-2002} : glass beads, sea dredged, crushed, and a mix
of $30 \%$ sea dredged and $70 \%$ crushed aggregates.
None of these coarse aggregates are monodisperse (the ratio of the
bigger particle size to the smaller one is close to $4$ for the all
the  aggregates).
Moreover, only the glass beads are spherical.
Then, most of the suspensions studied by \cite{Geiker-2002} do
not comply with the assumptions made to compute Eqs.~\refe{eq:taucKD}
and \refe{eq:etaKD}.

Consequently, it is not surprising that the data of
\cite{Erdogan-2005} are relatively well fitted by the theoretical
prediction while the fit between most of the experimental data
of~\cite{Geiker-2002} and the theoretical estimate is less
satisfactory.
More precisely, while it seems that Eq.~\refe{eq:etaKD} is able to
roughly estimate the consistency of suspensions of spherical particles
(glass beads) even if they are not monodisperse, 
the discrepancy between experimental and theoretical
data is no more acceptable for the other aggregates.
As depicted in Fig.~\ref{fig:viscosite-biblio}, tuning the value of the
maximum solid fraction does not allow to improve the quality of the
fit.

It is believed that this discrepancy comes from the fact that the sea
dredged, crushed and mixed aggregates are not spherical (an aspect
ratio equal to 2 has been measured for all these particles).
As the Krieger-Dougherty Eq.~\refe{eq:K-D} applies
only to suspensions of spherical particles, Eq. \refe{eq:etaKD} cannot
fit the properties of suspensions of non-spherical particles.
One concludes that using a function $g(\varphi)$
able to accurately estimate the linear properties of a suspension of
non spherical particles into Eqs.~\refe{eq:tauchom} and
\refe{eq:etahom} will improve the quality of the estimates. 
This opinion is supported by the fact \cite{Geiker-2002} observed
that the relative consistency of non-spherical particles suspensions
is greater than that of the glass beads suspensions when measured for
the same solid volume fraction (see Fig.~\ref{fig:viscosite-biblio}),
a trend classically observed for Newtonian suspensions.

\section{Validity of the theoretical model}
\begin{figure}[hbtp] 
\begin{center}
\input{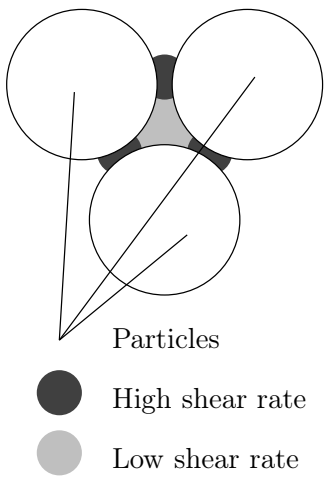}
\caption{Heterogeneities of the strain rate field for a concentrated
  suspension.
\label{fig:fluideetparticules}}
\end{center} 
\end{figure}

For all the results, the experimental
data are more dispersed when the solid volume fraction increases and the
discrepancy between the experimental and the theoretical results
increases when the solid volume fraction tends towards that of the maximum
packing fraction.
It could be  asserted that this discrepancy comes simply from the
dispersion of the experimental results.
Nevertheless, it is believed that heterogeneities of the secant modulus over
the liquid domain are not any more negligible for the larger values of
$\varphi$.
When the solid volume fraction value tends towards the maximum packing
fraction, the fluid located near the closest points of two adjacent
particles experiences a much larger shear rate than the fluid
located far from these points (see
Fig.~\ref{fig:fluideetparticules}).
In this situation, it is  necessary to take into account the
heterogeneities of the secant modulus in the framework of more
complex models in order to obtain more accurate estimates.

To use an uniform estimate of the secant modulus 
induces a  second drawback of the model.
As no difference is made between all the points located in the fluid
phase, such a model is unable to account for the fact that the fluid
 does not start to flow everywhere at the same time (or more precisely
that the yield stress is not reached everywhere at the same time in the
fluid domain).

For the Bingham suspension, \cite{Ponte-Castaneda-2003} has shown that beyond
the yield stress, the behavior of the suspension is first nonlinear and
then asymptotes to a purely linear behavior for higher strain rate values.
As this phenomenon is induced by the presence of strong fluctuations
of the strain rate field in the fluid phase at the onset of yield
which induce fluctuations of the secant modulus, the
simplified model described above is not able to account for this
nonlinearity.
Nevertheless, it is possible to show that the simplified approach performed
in this paper furnishes useful estimates.

Thanks to the properties of the minimization
problem~\refe{eq:LdCMacro2} it is  easily established that the
potential $W \left(\D\right)$ admits the following lower bound:
\begin{equation}
  \label{eq:Borne-W}
W\left(\D\right) \geq 
\Phi_1 (\D) + \Phi_2(\D)
\end{equation}
with 
\begin{equation}
  \label{eq:Borne-W-1}
  \Phi_1 \left(\D\right) = \left(1-\varphi\right) \tau_c  \min_{\d \in {\cal
        G}\left(\D\right)}\left\langle{\tilde d} \right\rangle_{\ell}
\end{equation}
and 
\begin{equation}
  \label{eq:Borne-W-2}
  \Phi_2 \left(\D\right) =  \left(1-\varphi\right) \frac{\eta}{n+1}
   \min_{\d \in {\cal G}\left(\D\right)} \left\langle{\tilde d}^{n+1}
   \right\rangle_{\ell}
\end{equation}
The two quantities $\Phi_1$ and $\Phi_2$ define the effective behavior
of two suspensions with the same microstructure as the yield stress
suspension.

$\Phi_1$ defines the constitutive law of a suspension of particles
embedded in a plastic material obeying the standard flow rule for a
von Mises criterion with yield  shear stress $\tau_c$.
$\Phi_2$ is the overall potential of a suspension of rigid particles
dispersed in a fluid whose behavior is described by the incompressible
viscous power-law:
\begin{equation}
  \label{eq:viscouslaw}
  \cauchy = 2 \eta {\tilde d}^{n-1} \d - p \tun \qquad \tr{\d} = 0
\end{equation}
Thanks to the properties of the function $\left\langle{\tilde d}^{m}
\right\rangle_{\ell}$ for $m \geq 1$ and of the set ${\cal
  G}\left(\D\right)$, it can be shown that the material defined by the
potential $\Phi_1$ is an isotropic rigid plastic material and the
overall potential $\Phi_2$ is that of an isotropic incompressible
viscous material whose overall behavior is described by an homogeneous
function of degree~$n$. 
Moreover, if the suspending fluid obeys the Bingham law ($n = 1$), the
effective potential $\Phi_2$ is that of a Newtonian fluid.

Then, we can state that the macroscopic behavior of
the isotropic suspension is energetically "underestimated" by a
yield stress state law with same exponent as that of the bearing
fluid.
Both the overall yield criterion and the viscous behavior can be
determined through the solution of two homogenization problems allowing
to explicitly calculate the lower bound~\refe{eq:Borne-W}. 
Of course, the overall potential of the yield stress suspension is not
the sum of two simple potentials. 
Nevertheless, it is obvious that the first term of the lower
bound~\refe{eq:Borne-W} prevails for the lower values of the
macroscopic strain rate $\D$, while for the larger values of the
strain rate, the second term must be taken into account. 
This property allows to show that, although the macroscopic behavior
of the suspension is not Herschel-Bulkley, it tends toward that of a
yield stress fluid with yield criterion defined by $\Phi_1$ and
viscous flow rule described by $\Phi_2$ for respectively the lower
values of the macroscopic strain rate (near the yield stress) and  the
larger values of the macroscopic strain rate (far from the yield
stress).
 \begin{figure}[htbp] 
\begin{center}
\input{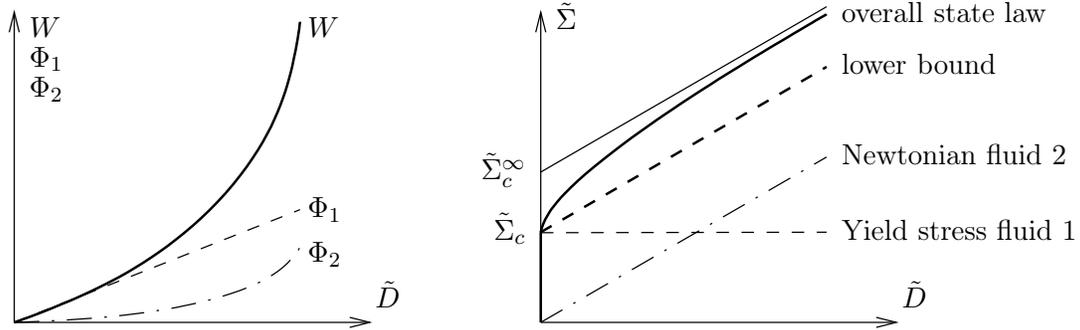}
\caption{One-dimensional sketch of the functions $W$, $\Phi_1$ and
  $\Phi_2$ as a function of the second invariant of
  $\D$ (left). The associated state laws (fictitious materials, lower
  bound and overall material) are depicted on the right part of the
  figure.
  The suspending fluid is Bingham ($n = 1$). 
\label{fig:lower-bound}}
\end{center} 
\end{figure}
These results are summarized in Fig.~\ref{fig:lower-bound} for a
suspension of particles in a Bingham fluid.
On the left part of Fig.~\ref{fig:lower-bound} are drawn the overall
potential $W$ and both potentials $\Phi_1$ and $\Phi_2$ as a function
of the second invariant of the macroscopic strain rate. 
For clarity, the lower bound $\Phi_1 + \Phi_2$ is not drawn. 
The overall behaviors of the suspension and of the materials
associated with the potentials $\Phi_1$, $\Phi_2$ and the lower bound
$\Phi_1 + \Phi_2$ are depicted on the right part of
Fig.~\ref{fig:lower-bound}.
This diagram is very similar to the Fig.~3 of
\cite{Ponte-Castaneda-2003}.
${\tilde \Sigma}$ denotes the second invariant of the macroscopic
deviatoric stress tensor. 

Interestingly, using the method described in
section~\ref{sec:secant}, it can be shown that Eqs~\refe{eq:taucKD} and
\refe{eq:etaKD} allow to estimate the value of the yield stress
defined by $\Phi_1$ and the value of the consistency defined by
$\Phi_2$.
Then, even if the overall behavior of the suspension is not exactly
Herschel-Bulkley, the estimates~\refe{eq:taucKD} and
\refe{eq:etaKD} can account for well defined properties of the
suspension.
Nevertheless it must be kept in mind that these estimates are not
associated to lower bounds of the potentials $\Phi_1$ and $\Phi_2$. 
A more accurate estimate for the overall behavior of a two-dimensional
Bingham suspension, which is of the form suggested by the solid curve
in the right part of Fig.~\ref{fig:lower-bound}, has been proposed
by~\cite{Ponte-Castaneda-2003}.

It is clear from Fig.~\ref{fig:lower-bound} that it
is not possible to deduce the value of the macroscopic yield stress
${\tilde \Sigma}_c$ from the measurements of the hight limit strain
rate behavior of the overall material.
Indeed, if one tries to evaluate the overall yield stress of the
suspension from  the high-limit strain rate behavior as
depicted in Fig~\ref{fig:lower-bound}, the obtained estimate, denoted
${\tilde \Sigma}_{c}^{\infty}$, is higher than the true yield stress
${\tilde \Sigma}_{c}$ one must apply to make the suspension flowing.
It seems that~\citet{Geiker-2002} used such a procedure to measure the
overall yield stress of the concretes they studied.
This result could explain why they measured
yield stress values much higher (ten times) than the ones measured by
others authors.

The theory predicts that for the higher values of the macroscopic
shear rate, the particles do not change the power-law index~$n$.
This result is in agreement with previous theoretical and experimental
approaches to the rheological properties of concentrated suspensions in
non-Newtonian fluid~[\citet{Polinski-1988,Ponte-Castaneda-2003}].  

\section{Conclusion}

Theoretical estimates for the overall rheological properties of a
suspension of noncolloidal and non-Brownian particles immersed in a
nonlinear fluid have been proposed. 
To our knowledge, this approach and the associated estimates are
original in the field of rheology. 
As the problem to be solved is nonlinear, it was necessary to make some
approximations in order to compute simple analytical estimates. 
Here, the estimates are valid provided that the heterogeneities of the
secant modulus can be neglected over the domain filled by the fluid
phase. 
Then, the overall properties of the nonlinear suspension are estimated from
that of a fictitious linear suspension having the same microstructure.

In the case where the suspending fluid obeys a Herschel-Bulkley law,
the estimates for the yield stress and the consistency read
$\tauchom/\tauc = \sqrt{(1-\varphi) g(\varphi)}$ and
$\etahom/\eta = \sqrt{g(\varphi)^{n+1}/(1-\varphi)^{n-1}}$ where
$g(\varphi)$ denotes the ratio of the macroscopic to the
microscopic properties of the fictitious linear material.  

Comparison of the theoretical results with experimental data for
suspensions of monodisperse rigid noncolloidal particles
dispersed in yield stress fluids have also been presented.
The experimental procedure used to obtained these data was designed
by~\cite{Mahaut-Chateau-Coussot-Ovarlez-2007} to evaluate the purely
mechanical contribution of the particles to the paste behavior, independently
of the physicochemical properties of the materials.

It was found from both the theoretical and the experimental approaches
that the dimensionless yield stress and the dimensionless consistency
depend on the bead volume fraction only. 
It has been shown that the yield stress/solid volume fraction and the
consistency/solid volume fraction relationships are well fitted to the
very simple laws $\tauchom/\tauc =
(1-\varphi)^{1/2}(1-\varphi/\varphi_m)^{-1.25 \varphi_m}$ and
$\etahom/\eta = \left(1-\varphi\right)^{(1-n)/{2}}
\left(1-\varphi/\varphi_m\right)^{-1.25(n+1) \varphi_m}$ respectively.

As long as the coarse particles are spherical, the theoretical
estimates agree quite well with experimental data found in the
literature even if experimental procedures used by the authors do not
exactly comply with the assumptions made to obtain the theoretical
results.
This is all the more satisfactory that only four numbers
are necessary to calculate the estimates of the overall properties of
the suspension as a function of the solid volume fraction: the close
packing density of the particles and three suspending fluid
properties (yield stress, consistency and power law index).

We now plan to study the case of polydisperse systems and that of
anisotropic particle distributions. 
To improve the theoretical estimates, it will also be necessary to separate
the fluid domain in several geometrical domains in order to better
describe the heterogeneities over the
liquid domain. 
In particular this situation is expected to be founded for
higher values of the solid volume fraction or for polydisperse
suspensions.

\bibliography{/home/xavier/bibtex/liste}

\end{document}